# Termination of two-dimensional metallic conduction near metal-insulator transition in a Si/SiGe quantum well

T.M. Lu,[1,*] W. Pan,[2] D.C. Tsui,[1] P.C. Liu,[3] Z. Zhang,[3] and Y.H. Xie[3]

[1] Department of Electrical Engineering, Princeton University, Princeton, New Jersey 08544, USA

[2] Sandia National Laboratories, Albuquerque, New Mexico 87185, USA

[3] Department of Materials Science and Engineering, University of California at Los Angeles, Los Angeles, California 90095, USA

Abstract

We report in this Letter our recent low temperature transport results in a Si/SiGe quantum well with moderate peak mobility. An apparent metal-insulating transition is observed. Within a small range of densities near the transition, the conductivity $\sigma$ displays non-monotonic temperature dependence. After an initial decrease at high temperatures, $\sigma$ first increases with decreasing temperature $T$, showing a metallic behavior. As $T$ continues decreasing, a downturn in $\sigma$ is observed. This downturn shifts to a lower $T$ at higher densities. More interestingly, the downturn temperature shows a power law dependence on the mobility at the downturn position, suggesting that a similar downturn is also expected to occur deep in the apparent metallic regime at albeit experimentally inaccessible temperatures. This thus hints that the observed metallic phase in 2D systems might be a finite temperature effect.

The physical properties of two-dimensional (2D) electrons have been a subject of interest for many years. Yet after years of research, the ground states of a 2D electron system (2DES) in the presence of disorder and electron-electron interaction, a realistic situation in experiments, remain an open question. Based on scaling arguments, Abrahams et al showed that an arbitrary amount of disorder is sufficient to localize all non-interacting 2D electrons [1]. This conventional wisdom was challenged, especially in 1994, when Kravchenko et al. reported the observation of a sharp drop of resistivity (ρ) with decreasing temperature ($T$) in a high mobility silicon metal-oxide-semiconductor field-effect transistor (Si-MOSFET) and proposed a possible metal-insulator transition (MIT) in the strongly interacting regime, characterized by the condition of $r_s \gg 1$, where $r_s$ is defined as the ratio of the Coulomb energy to the Fermi energy $E_F$ of the 2DES [2]. Soon after the report by Kravchenko et al., similar phenomena, with differences in details, were observed in 2DESs in various semiconductor epitaxial heterostructures. [3-5]

The apparent violation of the scaling theory has led to much theoretical and experimental research aiming at understanding the metallic conduction and the MIT. Though to date no consensus has been reached [6], it has become clear that the interplay between electron-electron interaction and disorder plays an important role in the observed apparent MIT. In particular, recent observations of a downturn in conductivity at low temperatures in Si-MOSFETs [7,8] and 2D holes in GaAs [9-11] seem to suggest that disorder plays an important role in the phenomenon and at $T = 0$ 2DES may eventually be insulating.

In this Letter, we report a quantitative study of this downturn in conductivity in a Si/SiGe quantum well (QW) from $T = 10$ K down to dilution refrigerator temperatures. The peak mobility in this structure is $1.7 \times 10^4$ cm$^2$/Vs at electron density $n = 1.4 \times 10^{11}$ cm$^{-2}$ and, in spite of this low mobility, low-temperature transport measurements can be carried out to ~ $10^{10}$ cm$^{-2}$ range. This combination of relatively low mobility and density allows us to observe the downturn in conductivity and to probe the transport properties in the diffusive regime within experimentally accessible range of $T$. It was observed that the temperature at which the downturn occurs shows a power law dependence on the mobility (μ) at the downturn position, indicating that a similar downturn is also expected to occur in the apparent metallic regime at albeit inaccessible temperatures. This thus suggests that the observed metallic phase in 2D systems may be a finite temperature effect and, at $T = 0$, the ground state is insulating. We also present quantitative comparisons to the theory of interaction corrections by Zala et al. [12] and the scaling theory by Punnoose et al. [13]

The material used was a modulation-doped Si/SiGe QW grown by molecular-beam-epitaxy [14]. Ohmic contacts were formed by alloying AuSb into the heterostructure. The electron density n was tuned by biasing the front gate, which consists of a stack of Al$_2$O$_3$, Cr, and Au. Low-temperature transport experiments were performed using standard lock-in techniques with excitation current as low as 100 pA at frequency as low as 1 Hz. Data between $T = 0.3$ K and $T = 10$ K was taken in a $^3$He cryostat, and data at lower $T$ was acquired in a dilution refrigerator. Two samples were examined in four cool-downs. The results from these two samples are consistent with each other. In the following, detailed data from one sample will be reported.

Figure 1(a) displays $\sigma(T)$ of the 2DES from $T = 0.3$ K to $T = 10$ K in the $1\times 10^{11}$ cm$^{-2}$ regime, in which the ratio of the Coulomb energy to the Fermi energy is ~10. The Fermi temperature $T_F$ and the ballistic-diffusive crossover temperature $\hbar/\pi k_B \tau$, estimated from the conductivity at $T = 0.3$ K, are also marked in the plot. For the region where $T \ll \hbar/\pi k_B \tau$ the 2DES is in the diffusive regime, which constitutes the major part of the data. At first glance the transport characteristics across the MIT appear similar to what have been observed in other Si systems [2, 15]. At high $n$ ($n \geq 1.28\times 10^{11}$ cm$^{-2}$) $\sigma$ increases monotonically with decreasing $T$, while at low $n$ ($n \leq 0.89\times 10^{11}$ cm$^{-2}$) $d\sigma/dT > 0$. At intermediate $n$, $\sigma$ is non-monotonic, decreasing with decreasing $T$ at high $T$, reaching a minimum at a characteristic temperature $T_{E1}$, and increasing upon further cooling. The minimum shifts to higher $T$ at higher $n$ and weakens in strength, and eventually $\sigma(T)$ becomes monotonic.

Closer examination of the data, however, reveals several features worth careful consideration. We first notice that in a small range of $n$ at which $\sigma(T)$ is non-monotonic, an additional extremum is observed at low $T$ and the 2DES turns insulating ($d\sigma/dT > 0$) below a $n$-dependent downturn temperature $T_{E2}$. The second feature worth noting is that $\sigma$ shows an approximately linear dependence on $T$ at $T \sim T_F$ across the transition. Lastly, deep in the metallic side, $\sigma$ keeps increasing with decreasing $T$ in the diffusive regime, showing no sign of saturation. Each of these characteristics is discussed in the following.

The curves near the MIT are magnified and shown in Fig. 1(b), where $T_{E1}$ ($T_{E2}$) are marked by the upwards (downwards) arrows. For the curves shown in red, the 2DES exhibits a reentrant behavior upon cooling, being insulating above $T_{E1}$, metallic at intermediate $T$, and insulating below $T_{E2}$. At higher or lower n this behavior is not observable. The $n$ dependence can be seen more clearly in a phase diagram plot of $T$ and $n$, which is obtained from the sign of $d\sigma/dT$ and is shown in Fig. 2. Also shown in the diagram are $T_F$ and $T_D = \hbar/\pi k_B \tau$. As discussed above, $T_{E1}$ moves to higher $T$ at higher $n$, approximately following $T_F$. On the other hand, $T_{E2}$ shifts to lower $T$ as n is increased, similar to $\hbar/\pi k_B \tau$. Extrapolating this trend to higher $n$ would then imply an insulating behavior at low enough $T$.

We have also examined $\sigma(T)$ in the same specimen down to dilution refrigerator temperatures. $T_{E2}$ in this cool down shifts to lower $T$ with increasing $n$. In Fig.3a, $T_{E2}$ from the two sets of data is plotted as a function of $n$. Overall the trend that $T_{E2}$ shifts to a lower temperature at higher densities is the same for the two cool downs. The disagreement from the same sample in the two cool-downs is attributed to slightly different disorder configurations from one cool down to the other. In Fig.3b, $T_{E2}$ is plotted as a function of $\mu(T=T_{E2})$, the mobility where the downturn occurs. Surprisingly, the two sets of data collapse onto a single curve with $T_{E2}$ decreasing with increasing $\mu$ ($T=T_{E2}$), or in turn, $\tau$. A power law fit shows that $T_{E2}$ approximately scales as $T_{E2} \propto \mu^{-2.2}$.

The downturn behavior was observed in other Si/SiGe QW samples, for example, one with a similar growth structure and sample quality examined in this work and one reported in Ref. [5]. $T_{E2}$'s in these samples are also included in Fig. 3(b) and in the inset. We shall note here that, though the downturn occurs at a relatively high density (n=4.05×10$^{11}$ cm$^{-2}$) in Ref. [5], yet in the plot of $T_{E2}$ vs µ($T_{E2}$) their result is in fairly good agreement with the two other samples we looked at.

Similar decrease in conductivity with decreasing $T$ at low temperatures on the metallic side has also been observed in different material systems (for example, in Si-MOSFETs [7,8] and in dilute 2D hole systems in GaAs [9-11]), and was taken as evidence of the onset of weak-localization. However, we want to caution that in the experiments on 2D holes in GaAs and in one experiment on Si-MOSFET, the 2D conductivity is much larger than $e^2/h$, while in our case the downturn occurs in the regime where $\sigma \sim e^2/h$. Thus, it is unclear whether these qualitatively similar phenomena in different regimes are of the same origin.

We emphasize here that in the transition regime where $T_{E1}$ and $T_{E2}$ coexist the conductivity is of the order of $e^2/h$, indicative that both electron-electron and electron-disorder interactions are important. In the following, we propose to understand the existence of $T_{E2}$ under the picture of a recently proposed micro-emulsion model [6]. Within this model, the strongly interacting electron system consists of two components, Wigner crystallites dispersed into a Fermi liquid. The resistivity of 2DES is proportional to the viscosity of this micro-emulsion. For temperatures not too low, lowering temperature actually helps to reduce the percentage volume of the crystallites. This counter-intuitive behavior can be understood as follows. The Winger crystal state is determined by the free energy, $F = E-TS$, where $E$ is the energy at $T = 0$, and $S$ is the entropy. Lowering $T$ increases its free energy and helps to destabilize Wigner crystal state. In other words, the crystal melts as $T$ is lowered. Due to this reduction of crystallites, micro-emulsion becomes less viscous. The 2DES resistivity thus decreases, or its conductivity increases, upon cooling. As $T$ is further decreased, a counter transition to the melting of the crystallites can now occur and the percentage volume of the pinned Wigner solid increases. It can be imagined that these two processes may reach a balance at $T_{E2}$ when the conductivity reaches a maximal value. For $T < T_{E2}$, the process of forming a highly disordered Wigner solid prevails and the 2DES conductivity decreases with decreasing $T$. The above scenario can qualitatively explain several features on $T_{E2}$. First, in high mobility samples, there are less pinning centers. As a result, the balance point, or $T_{E2}$, has to be shifted to lower $T$, consistent with what we observed. Second, disorder potential varies from one sample to another. For example, in Si-MOSFETs electron wavefunction is close to the interface between Si and SiO$_2$ and the 2DES experiences mostly scattering by surface roughness and interface charges. In Si/SiGe QWs, the carriers are from the modulation doping and surface roughness scattering is reduced. The 2DES experiences mostly the disorder potential due to the modulation doping impurities and dislocation defects at the interface due to the Si and SiGe lattice mismatch. This difference in the nature of disorder is expected to affect the pinning strength and thus the power-law exponent. Indeed, a power-law dependence of $T_{E2}$ also exists in data obtained in the high quality Si-MOSFET [7], with a power law exponent of ~1, different from our results.

At the present time, our data does not allow us to definitively pin down the nature of the insulating phase at $T = 0$. In the paper by Prus et al [7], the downturn was attributed to both the weak-localization interference and Coulomb interaction effects. In our experiment, the conductivity downturn occurs at lower densities in the transition region where the weak-localization was shown to be strongly suppressed [16]. We thus propose that the insulating phase in our sample is of strong electron-electron interaction origin. One such candidate, for example, is the insulating Wigner glass state which has been proposed as a $T = 0$ ground state at intermediate carrier concentration in the presence of disorder [17]. Of course, more experiments are needed to verify whether this Wigner glass phase is responsible for the observed insulating behavior for $T < T_{E2}$.

We now turn our attention to $\sigma(T)$ deep in the metallic regime where $\sigma \gg e^2/h$. At the highest n, $\sigma$ increases with decreasing T and does not show sign of saturation even when $T \ll \hbar/\pi k_B \tau$, as shown in Fig. 1(a). In Fig. 4(a) and (b) we show $\sigma(T)$ at $n = 1.4 \times 10^{11}$ cm$^{-2}$ in linear-$T$ and log-$T$ scales, respectively. At high $T$ ($T > 0.3$ K), $\sigma$ increases linearly with decreasing $T$, as shown in Fig. 4(a). Upon further cooling, $\sigma$ increases even faster than the linear dependence and approximately follows a log-$T$ dependence, as shown in Fig. 4(b). Such strongly enhanced metallic behavior in the diffusive regime is not observed in Si-MOSFETs, which show saturation of $\sigma$ at dilution refrigerator temperatures [18]. It was shown by Zala et al. that in the ballistic regime, the correction is linear in $T$, while in the diffusive regime the correction is logarithmic [12]. The sign and magnitude of the correction depends on the strength of electron-electron interaction, characterized by the parameter $F^\sigma_0$. Extrapolating the linear portion of $\sigma(T)$ to $T = 0$ gives the conductivity from which we calculate the ballistic-diffusive crossover temperature, ~ 0.1 $\hbar/k_B\tau$ [19], marked by the green bar in Fig. 4. The value of $F^\sigma_0$ can be obtained from the slopes of $\sigma(T)$ at high $T$ and $\sigma(\log$-$T)$ at low $T$. The so determined values of $F^\sigma_0$ in the ballistic regime and the diffusive regime are -0.15 and -0.31, differing by a factor of two. Considering the contribution of weak localization, which also shows a logarithmic correction, makes this discrepancy even larger. We note that different $F^\sigma_0$ values in different regimes (i.e., diffusive vs ballistic) have also been reported before, for example, by Renard et al in Ref. [19].

Before we conclude this paper, two remarks are in order. First, near the MIT, $\sigma$ increases with $T$ in an approximately linear fashion at $T \sim T_F$. Different from previous studies, d$\sigma$/d$T$ in this sample displays non-monotonic density dependence with a broad peak and the peak value is ~ 0.03 $e^2/h$ per Kelvin at $n \sim 1 \times 10^{11}$ cm$^{-2}$. This peak value is much smaller than that observed in other high mobility 2DESs [20,21]. Currently there is no quantitative description of this regime where electrons strongly interact and $T \sim T_F \sim \hbar/\pi k_B \tau$. Our data suggests that while qualitatively similar phenomena are observed in both high mobility and low mobility 2DESs, disorder play an important role in a quantitative model.

Second, we have also compared our data with scaling theory by Punnoose et al. which was developed for the diffusive regime and shown to describe ρ(*T*) of 2DESs in Si-MOSFETs near the MIT quantitatively by a universal function without any adjustable parameters [13,22]. Our data, however, cannot be collapsed onto the universal function even if we focus only on the curves with the maximal $\rho < \pi h/e^2$, as required by the theory. We thus conclude that the universal function does not quantitatively describe our experimental data.

We thank L. Engel, Z. Jiang, H. Zhu, and P. Jiang for their help. The work at Princeton University was funded by the DOE under Grant No. DE-FG02-98ER45683 and the NSF under Grant No. DMR-0803730. The work at Sandia was supported by the DOE Office of Basic Energy Sciences. Sandia National Laboratories is a multi-program laboratory managed and operated by Sandia Corporation, a wholly owned subsidiary of Lockheed Martin company, for the U.S. Department of Energy's National Nuclear Security Administration under contract DE-AC04-94AL85000.

* Electronic mail: tmlu@princeton.edu

Figure Captions:

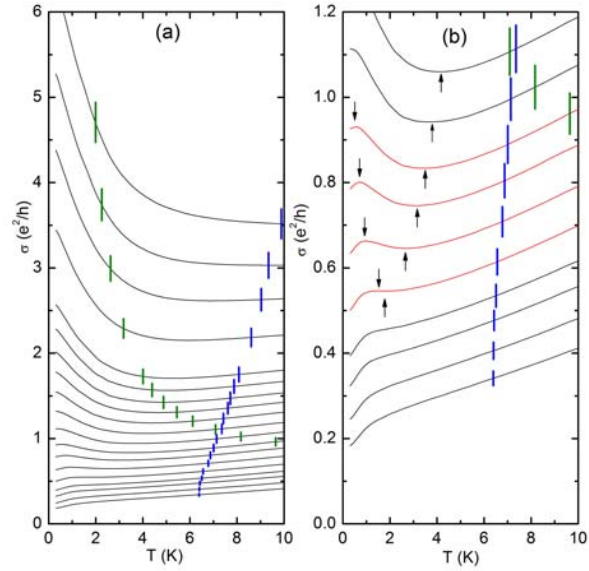

Figure 1 (Color) (a) σ in units of $e^2/h$ as a function of $T$ from 0.3 to 10 K. The electron densities from top to bottom are 1.35, 1.28, 1.23, 1.18, 1.11, 1.08, 1.06, 1.04, 1.015, 1.006, 0.976, 0.959, 0.040, 0.927, 0.897, 0.890, 0.878, 0.876, 0.874 × $10^{11}$ cm$^{-2}$. The blue (green) bars mark $T_F$ ($\hbar/\pi k_B \tau$). (b) Zoom-in of the curves with n ≤ 1.006×$10^{11}$ cm$^{-2}$ near the MIT region. $T_{E1}$ ($T_{E2}$) are marked by the upward (downward) arrows. Curves that show a downturn with decreasing T are displayed in red.

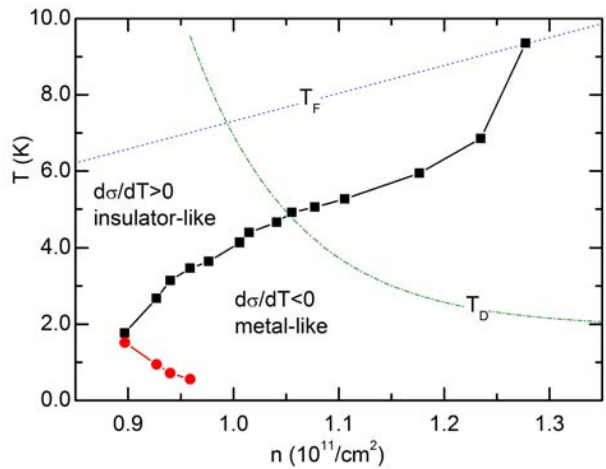

Figure 2 (Color online) A phase diagram of $T$ and $n$, based on the sign of dσ/d$T$. The black squares (red dots) represent $T_{E1}$ ($T_{E2}$). $T_F$ and $T_D = \hbar/\pi k_B \tau$ are shown as the dotted and dash-dotted lines, respectively.

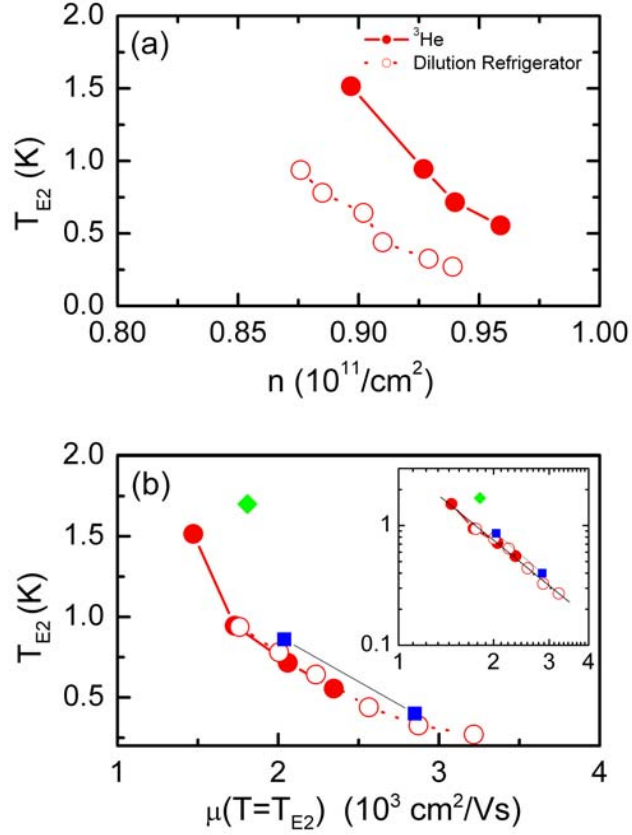

Figure 3 (Color online) (a) $T_{E2}$ as a function of $n$; (b) $T_{E2}$ as a function of $\mu(T = T_{E2})$. The inset shows the same plot and a power-law fit (dotted line) in logarithmic scale. Data from another two samples, one examined in this work (squares) and the other in Ref. [5] (diamond), are also included.

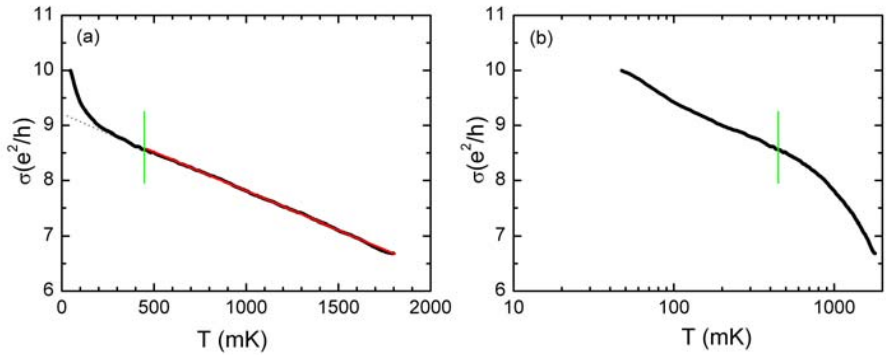

Figure 4 (Color online) $\sigma(T)$ at $n = 1.4 \times 10^{11}$ cm$^{-2}$ in (a) linear-$T$ and (b) log-$T$ scales. The green bars mark the crossover temperature, $0.1\ \hbar/k_B\tau$. The red line in (a) shows the linear section, and the dotted line shows the extrapolation to $T = 0$.